\documentclass[12pt]{article}
\usepackage{makeidx,color}

\usepackage{epsfig}
\usepackage{inputenc}
\usepackage{html}
\inputencoding{latin1}

\def\version{0.0}

\def\code#1{\texttt{\textbf{#1}}}
\def\urllink#1#2{\htmladdnormallink{\code{#1}}{#2}}
\def\url#1{\urllink{#1}{#1}}
\def\shell#1{\texttt{#1}}
\def\hclass#1{\urllink{#1}{../#1.html}}
\def\classh#1#2{\urllink{#1}{../#2.html}}
\def\event{\hclass{Event}}
\def\egen{\hclass{Event\-Generator}}
\def\fullgen{\hclass{Full\-Event\-Generator}}
\def\partgen{\hclass{Partial\-Event\-Generator}}
\def\collision{\hclass{Collision}}
\def\subprocess{\hclass{Sub\-Process}}
\def\step{\hclass{Step}}
\def\steph{\hclass{Step\-Handler}}
\def\eventh{\hclass{Event\-Handler}}
\def\collh{\hclass{Collision\-Handler}}
\def\particle{\hclass{Particle}}
\def\pdata{\hclass{Particle\-Data}}
\def\dmode{\hclass{Decay\-Mode}}
\def\decayer{\hclass{Decayer}}
\def\repository{\hclass{Repository}}
\def\repo{\hclass{Repository}}
\def\sm{\hclass{Standard\-Model\-Base}}
\def\rnd{\hclass{Random\-Generator}}
\def\strat{\hclass{Strategy}}
\def\lumifn{\hclass{Luminosity\-Function}}
\def\subh{\hclass{Sub\-Process\-Handler}}
\def\pextract{\hclass{Parton\-Extractor}}
\def\pxsec{\hclass{Parton\-XSecFn}}
\def\pdfbase{\hclass{PDF\-Base}}
\def\remh{\hclass{Remnant\-Handler}}
\def\partcoll{\hclass{Partial\-Collision\-Handler}}

\def\cascade{\hclass{Cascade\-Handler}}
\def\frag{\hclass{HadronizationHandler}}
\def\decayh{\hclass{Decay\-Handler}}
\def\anah{\hclass{Analysis\-Handler}}
\def\manip{\hclass{Event\-Manipulator}}
\def\interfaced{\hclass{Inter\-faced}}
\def\beamdata{\hclass{Beam\-Particle\-Data}}
\def\hint{\hclass{Hint}}
\def\lastinfo{\classh{Last\-XComb\-Info}{XComb}}
\def\lvec{\hclass{Lorentz\-5\-Vector}}
\def\kincut{\hclass{Kinematical\-Cuts}}
\def\String{\hclass{String}}
\def\lundfragh{\hclass{Lund\-Frag\-Handler}}
\def\lundptgen{\hclass{Lund\-Pt\-Generator}}
\def\lundzgen{\hclass{Lund\-Z\-Generator}}
\def\lundflavourgen{\hclass{Lund\-Flavour\-Generator}}
\def\endp{\hclass{EndPoint}}
\def\hadron{\hclass{Hadron}}

\def\clhep{CLHEP}
\def\SIUnits{SIunits}
\def\fortran{Fortran}
\def\linux{Linux}
\def\gnu{GNU}
\def\gcc{GCC}
\def\sub#1{_{\mbox{\scriptsize #1}}}
\def\m#1{\mbox{#1}}
\def\cpp{{\sc C++}}
\def\ariadne{{\sc Ariadne}}

\def\jetset{{\sc Jetset}}
\def\pythia{{\sc Pythia}}
\def\pyth{{\sc Pythia7}}

\def\xf{x\!f}

\def\fig#1{fig.~\ref{#1}}

\def\Caption#1{\caption[dummy]{{\it #1}}}
\def\laeq{\,\lower3pt\hbox{$\buildrel < \over\sim$}\,}

\def\sigh{\hat{\sigma}}
\def\shat{\hat{s}}
\def\that{\hat{t}}

\newcommand{\Publication}[4]{{\it #1} {\bf #2} ({#4}) {#3}}

\newcommand{\NPB}[3]{\Publication{Nucl.~Phys.}{B#1}{#2}{#3}}

\newcommand{\PR}[3]{\Publication{Phys.~Rep.}{#1}{#2}{#3}}
\newcommand{\CPC}[3]{\Publication{Comput.~Phys.~Comm.}{#1}{#2}{#3}}

\newcounter{Aenumct}

\renewcommand{\descriptionlabel}[1]%
{\code{#1}\hspace{-2mm}}

\begin{document}

\begin{titlepage}

  \renewcommand{\thefootnote}{\fnsymbol{footnote}}

  \begin{flushright}
    LU--TP 00--23\\
    hep-ph/0006152\\
    May 2000
  \end{flushright}
  \begin{center}
    
    \vskip 10mm
    \textbf{\LARGE \pythia\ version 7-\version\\[4mm]
      -- a proof-of-concept version\footnote{Work supported by the EU
        Fourth Framework Programme `Training and Mobility of
        Researchers', Network `Quantum Chromodynamics and the Deep
        Structure of Elementary Particles', contract FMRX-CT98-0194
        (DG 12 - MIHT).}}
    \vskip 15mm
    {\large Marc Bertini, Leif Lönnblad and Torbjörn Sjöstrand}\\
    Department of Theoretical Physics\\
    Lund University\\
    Sölvegatan 14A\\
    S-223 62  Lund, Sweden\\
    marc@thep.lu.se, leif@thep.lu.se, torbjorn@thep.lu.se

  \end{center}
  \vskip 50mm
  \begin{abstract}

    \hskip -3mm
    
    This document describes the first proof-of-concept
    version of the \pyth\ program.
    
    \pyth\ is a complete re-write of the \pythia\ program in \cpp. It
    is mainly intended to be a replacement for the `Lund' family of
    event generators, but is also a toolkit with a structure suitable
    for implementing any event generator model.
    
    In this document, the structure of the program is presented both
    from the user and the developer point of view. It is not intended to
    be a complete manual, but together with the documentation provided
    in the distribution, it should be sufficient to start working with
    the program.

  \end{abstract}

\end{titlepage}

\def\contentsline#1{\vskip -2mm\csname l@#1\endcsname}
\tableofcontents

\newpage

\section{Introduction}
\label{sec:intro}

\pyth\ \cite{pyt7} will be a new event generator well suited to meet
the needs of future high-energy physics projects, for phenomenological
and experimental studies. The main target is the LHC community, but it
will work equally well for linear e$^+$e$^-$ colliders, muon
colliders, the upgraded Tevatron, and so on.  The generator will be
based on the existing Lund program family, but rewritten from scratch
in a modern, object-oriented style, using \cpp. The greatly enhanced
structure will make for improved ease of use, extendibility to new
physics aspects, and compatibility with other software likely to be
used in the future.

\subsection{Motivation}
\label{sec:intro:motivation}

The current state-of-the-art event generator programs from the Lund
group --- \pythia, \jetset\ and \ariadne\ --- generally work well. It
has been possible gradually to extend them well beyond what could
originally have been foreseen, and thus to parallel the development of
the high-energy physics field as a whole towards ever more complex
analyses. However, a limit is now being approached, where a radical
revision is necessary, both of the underlying structure and of the
user interface.

Even more importantly, there is a change of programming paradigm,
towards object-oriented methodology. In the past, particle physicists
have used Fortran, but now \cpp\ is taking over. \cpp\ has been
adopted by CERN as the main language for the LHC era. The CERN program
library is partly going to be replaced by commercial products, partly
be rewritten in \cpp.  In particular, the rewriting of the detector
simulation program Geant \cite{Geant4} is a major ongoing project,
involving several full-time programmers for many years, plus voluntary
efforts from a larger community.  The CERN shift is matched by
corresponding decisions elsewhere in the world: at SLAC for the
B-factory, at Fermilab for the Tevatron Run 2, and so on.

Therefore a completely new version of the Lund programs, written in
\cpp, is urgently called for. What is required is a complete
rethinking of the way an event generator should look.

Some attempts have already been made. Although the MC++
program\cite{MCPP} never left the toy stage, pieces of it still live
on, since a subset of the classes became the foundation of
\clhep\cite{CLHEP}. Lately there has been some programs written for
parts of the event generation process, e.g.\ 
APACIC/AMEGIC\cite{amegic}, and a couple of attempts to implement a
general event record in \cpp\cite{hepmc,stdheppp}, but \pyth\ is the
first serious attempt to write a complete event generator in \cpp.

\subsection{Strategy}
\label{sec:intro:strategy}

The main idea is to define a structure which encapsulates the event
generation process. The standard way of generating events is first to
choose a hard scattering sub-process from parton densities and hard
parton-parton matrix elements. The partons in the scattering are then
allowed to develop perturbative showers, and the resulting partons are
hadronized. Finally the produced hadrons are decayed until only stable
particles remain. There is normally a natural ordering of these steps
given by the physical scale involved in the corresponding processes,
but the order in which different steps are performed is not always
straightforward, and the event generation process can become very
complicated e.g. when multiple interactions are included.

In \pyth, all the different steps are described in terms of a set of
abstract base classes with well defined interfaces using virtual
member functions. Any model for a specific step in the event
generating process can then be implemented in a class which inherits
from the appropriate base class. In addition, any such model can be
modified by further inheritance. There are several such base classes
in \pyth, some of which are shown below in \fig{fig:handlers}.

A number of models implemented like this can then at run-time be
combined into an \egen\ which can produce \event s. An \event\ is, of
course, also an object representing the produced particles. But the
\event\ class is not just a list of particle entries, as the
\textit{event records} of current \fortran-based generators. It is a
highly structured class containing the complete history of the
generation of each event. Of course, if the user is only interested in
a list of final state particles, this is also easily extracted.

\subsection{About this document}
\label{sec:intro:about}

The layout of this paper is as follows. Section \ref{sec:desc}
describes how to obtain, install and run a sample \pyth\ program.
Section \ref{sec:user} contains a more detailed description of how to
use the program, while section \ref{sec:develop} describes how to
implement models in the \pyth\ framework. In section
\ref{sec:lundfrag}, the implementation of Lund string fragmentation is
given as a case study. Finally in section \ref {sec:bugs} there is a
description of things presently lacking in the program and of bugs
which have not yet been corrected.

This is not intended to be a complete manual, but in the HTML version
of this paper, which is included in the program distribution, all
classes mentioned are linked to html-ized header files, and together
they should provide a reasonable guide for anyone who wants to use the
program and/or implement physical models.

\subsection{Future plans}
\label{sec:intro:future}

It should be noted that the program described here is not of
production quality and results produced with it should not be used for
serious scientific studies. This release is mainly intended to make
public the basic structure of the program.

The work to implement real physical models into this framework has
started. String fragmentation is e.g.\ already present (see section
\ref{sec:lundfrag} and ref.\ \cite{Marc}).  The plan is to have
something of production quality within a year from now, although it
may take several years before it can fully make obsolete the current
\fortran\ version of \pythia.

\section{Description}
\label{sec:desc}

To run the current \pyth\ program, a computer running \linux\ and the
\gnu\ \gcc\ compiler version 2.95.2 is needed. The aim is to have
\pyth\ written fully according to the ANSI/ISO \cpp\ standard, so in
principle it should be possible to run it on any computer with a
standard-compliant compiler. However, at present there are very few
(if any) fully standard compliant compilers available and, as a
consequence, inadvertently, there may be some non-standard code in
\pyth. To avoid problems, it is therefore recommended to use \pyth\ 
with the \gcc\ compiler.

To run \pyth\ you also need to have the \clhep\ library installed,
which is available from
\url{http://wwwinfo.cern.ch/asd/lhc++/index.html}.

\subsection{Installation}
\label{sec:desc:install}

The program is available on the web at\\
\urllink{http://www.thep.lu.se/Pythia7/Pythia7-\version.tar.gz}{http://www.thep.lu.se/Pythia7/Pythia7-\version.tar.gz}.\\
Unpack the distribution on a computer running \linux\ by doing\\
\shell{gunzip -c pythia7-\version.tar.gz | tar xf -}\\
This will produce a directory called \shell{Pythia7-\version/Pythia7}.

To compile, you first need to run the \shell{configure} script in this
directory. This will set up the \textit{Makefile}s to use \shell{g++}
as compiler and to look for the \clhep\ include and library files in
the \shell{include} and \shell{lib} directories under
\shell{/usr/local}. These choices can be changed by setting the shell
environment variables \shell{CXX} and \shell{CLHEPPATH} to the desired
\cpp\ compiler and to the path where \clhep\ is installed
respectively. This should be done before the configure script is run.
After this, \shell{make current} will compile the main \pyth\ library
file and a couple of sample programs in the \shell{src} subdirectory.

\subsection{Running the sample generator}
\label{sec:desc:run}

There are three sample programs in the \shell{src} subdirectory:
\begin{itemize}
\item \shell{setupPythia SimpleLEP.in} reads in the default
  \emph{repository} of objects and the commands in
  \shell{SimpleLEP.in} to setup a simple \egen\ object capable of
  generating e$^+$e$^-\!\!\rightarrow q\bar{q}$ annihilation events at
  91~GeV. This generator is then written to a file called
  \shell{SimpleLEP.run}.
\item \shell{runPythia SimpleLEP.run} reads the generator and runs it.
  This will produce three output files: \shell{SimpleLEP.log} will
  contain a printout of the first ten events and possible error
  messages; \shell{SimpleLEP.out} will contain the results of the run
  -- in this case only the integrated cross section; and
  \shell{SimpleLEP.tex} will contain a description of the models used
  in the run including the relevant references.
\item Finally, there is the \shell{runPartial.cc} program which is an
  example of how to use \pyth\ in other applications where full events
  are not required. \shell{runPartial SimpleLEP.run} will simply
  hadronize a simple partonic system specified in the program.
\end{itemize}

The \shell{setupPythia} and \shell{runPythia} programs will be run
automatically with \shell{make check}.

\section{User information}
\label{sec:user}

The normal way to use \pyth\ is to run a setup program to define an
\egen. This generator can then be used in an application to generate
events to be used there. But it is also possible to setup a \fullgen,
so that the full run, including analysis, can be defined. In this case
the \shell{runPythia} program can be used to do everything.

\subsection{The event record}
\label{sec:user:event}

\begin{figure}[htbp]
  \begin{center}
    \epsfig{figure=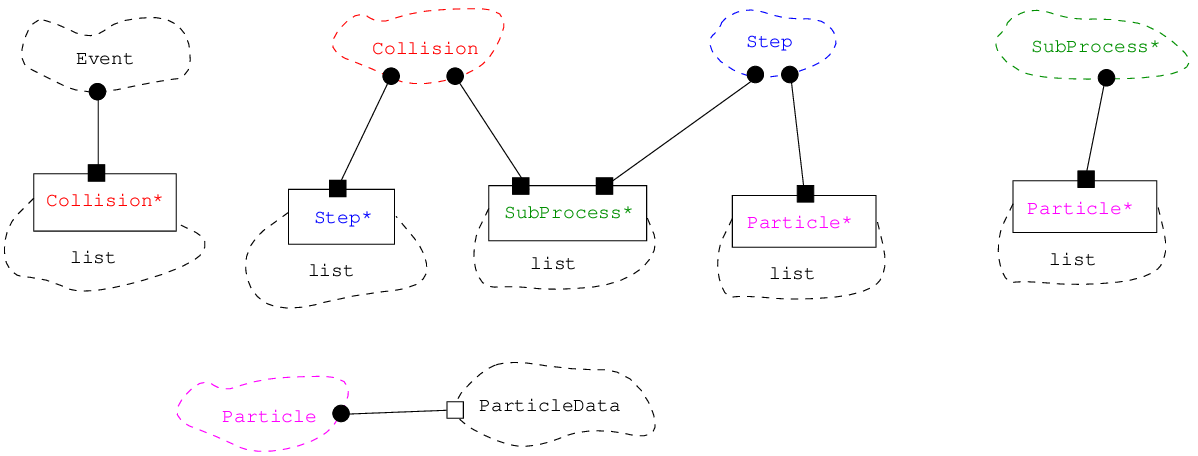,width=15cm}
  \end{center}
  \Caption{Class diagram for the structure of a generated event in \pyth.}
  \label{fig:event}
\end{figure}

To inspect and analyze a generated event the user is given an \event\ 
object which contains a very detailed account of how the event was
generated. The structure of the event classes is shown in
\fig{fig:event}. At high luminosity hadron colliders an event usually
consists of several collisions and consequently the \event\ object has
a list of \collision\ objects. Each \collision\ has a list of
\subprocess\ objects, one primary \subprocess\ and, in case of multiple
interactions, a number of secondary ones. The \collision\ has also a
list of \step\ objects, each of which has a list of \particle s
corresponding to the full state of the event after a given step in the
event generation process. The last one of these represents the
final-state particles which may be detected.

This is a fairly complicated structure, to allow for detailed analysis
of the generation. But the normal user need not worry about the inner
structure of an \event. Instead there are a number of member functions
which can be used to extract the information needed. The main such
method is called \code{select}. Given a so-called predicate object, it
will extract particles which satisfy the conditions specified in the
predicate into a container. The only requirement on the predicate
object is that it is derived from the \hclass{SelectorBase} class.
\pyth\ will contain a number of such selector classes, the most common
one being the \code{SelectFinalState} which simply extracts a list of
all final state particles.

The \particle\ class is fairly straightforward. It contains
information about an instance of a particle, such as its momentum and
creation point, while all properties of the corresponding particle
type can be accessed through a pointer to a \pdata\ object.

The \particle\ object also contains a list of (pointers to) its
parents and, in case it has decayed, a list of children. In case the
particle is coloured, it also contains pointers to its (anti-) colour
neighbours and pointers to the parents and children from/to which it
has obtained/given its (anti-) colour.

In some cases the state of a particle instance may have been changed
in a generation step, even if it has not decayed, e.g.\ its energy may
have been modified in order to conserve the total energy in some
process. For this reason, a particle may have a pointer to objects
corresponding to the previous and/or next instance of the same
particle.

\subsection{Particle properties}
\label{sec:user:particle}

The \pdata\ objects contain all information about a particle type: its
id-number (according to the PDG standard\cite{PDGID}), its name (for
which there is no standard yet), its nominal mass and width, its
charge, spin and colour quantum numbers, etc.

\pdata\ also has a list of \dmode s, each of which, in the simplest
form, contains a branching fraction, a list of \pdata\ objects
corresponding to the decay products and a pointer to a \decayer\ 
object which is capable of performing the decay. But a \dmode\ may
also represent more abstract decay modes such as
W$^+\!\!\rightarrow$\textit{hadrons} or
B$^+\!\!\rightarrow\mu^+\nu_\mu+$\textit{anything}, and also decay
chains such as
H$^0\!\!\rightarrow$W$^+$W$^-\!\!\rightarrow$e$^+\nu\sub{e}\bar{\m{u}}$d.
This is achieved by the introduction of \emph{particle matcher}
objects (of base class \hclass{MatcherBase}), which represent a whole
set of particles, and by specifying \dmode s of the decay products.

The \pdata\ object may also have a pointer to a \hclass{MassGenerator}
object which can generate the mass of a given particle instance, given
the nominal mass and width. Also a width generator\footnote{Not yet
  implemented.} may be assigned to a \pdata\ object to calculate the
width and partial widths of a particle according to some (e.g.\ SUSY)
model.

A particle may have different kinds of masses. The mass given in the
\pdata\ class corresponds to the kinematical mass, i.e.\ 
$\sqrt{E^2-p^2}$, of a real particle. Quarks and di-quarks may be
given as objects of the \hclass{Constituent\-Particle\-Data} class
which inherits from \pdata\ and carries information about the
\emph{constituent} mass. Currently there is no mechanism for
specifying other kinds of masses for particles, e.g.\ scale-dependent
masses, but this will be possible in the future.

\subsection{Setting up an event generator run}
\label{sec:user:setup}

To setup an event generator run, one has to manipulate the \repository\ 
which has a list of all objects available for use in \pyth. The idea is
then to connect different objects with each other (e.g.\ assign a
\hclass{MassGenerator} to a \pdata\ object above). Through the
\repository\ it is also possible to modify parameters and switches
which may be available in the objects (e.g.\ the nominal mass in a
\pdata\ object).

\subsubsection{The handler classes}
\label{sec:user:handler}

\begin{figure}[htbp]
  \begin{center}
    \epsfig{figure=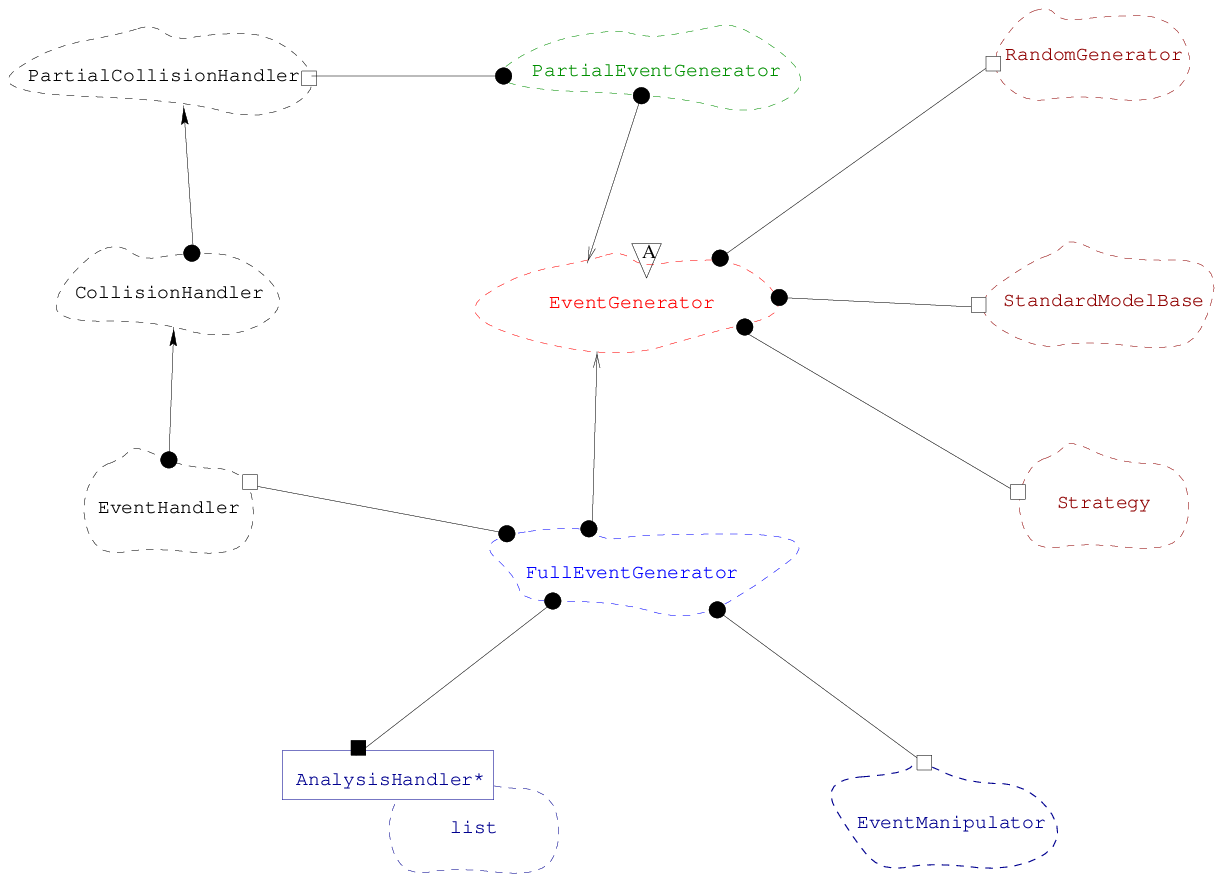,width=13cm}
  \end{center}

  \begin{center}
    \epsfig{figure=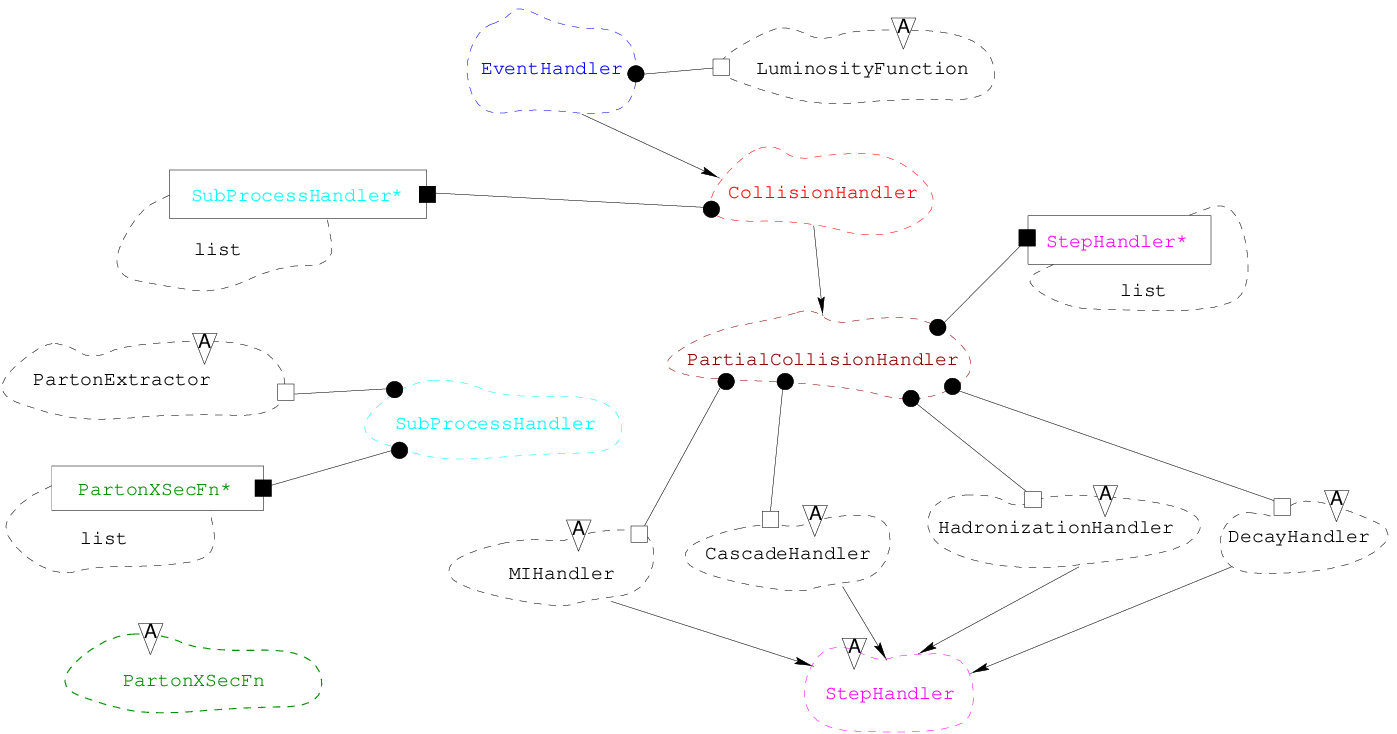,width=15cm}
  \end{center}
  \Caption{Class diagram for the structure of different handlers
    responsible for the generation process in \pyth.}
  \label{fig:handlers}
\end{figure}

The \repository\ will by default contain a large number of sample
\egen s, so the ordinary user should never have to setup an \egen\ 
from scratch. In \fig{fig:handlers} is shown the structure of the main
handler classes in \pyth. As seen, there are two \egen\ sub-classes,
the \fullgen\ and the \partgen. The latter is a scaled down version
which is not able to generate full collisions but can be used inside
other applications to apply \steph s to a user-supplied initial \step.

Most importantly, the \fullgen\ has a pointer to an \eventh\ which
performs the actual generation of an event. In addition the \egen\ 
contains objects which are global to a particular run, such as a \sm\ 
object which implements the Standard Model parameters to be used, a
\rnd\ object, the list of \pdata\ objects to be used, etc. Some of
these global objects may be collected in a \strat\ object.

\subsubsection{Sub-process selection}
\label{sec:user:handler:subp}

The \eventh\ uses a \lumifn\ to obtain a pair of colliding
beam-particles which are then used by the inherited \collh\ to perform
the actual collision. The \collh\ may have several \subh s to manage
different kinds of sub-processes, each of which has a \pextract\ 
object and a list of \pxsec\ objects. The \pextract\ handles the
extraction of partons from the beam-particles and the generation
of remnants, while the \pxsec\ implements hard parton--parton
scatterings. The \pextract\ can be assigned parton density objects of
class \pdfbase, but normally the colliding particles are given as
\beamdata\ objects, in which case a \pdfbase\ object is specified
there. To each \pdfbase\ there is assigned a \remh\ object which is
used by the \pextract\ to generate the remnants.

The selection of a hard sub-collision is a two-step procedure. The
\eventh\ has a list of so-called \hclass{XComb}s, one for each possible
combination of \subh, \pxsec\ and pair of incoming partons. Given a
\kincut\ object, an \hclass{XComb} is chosen according to an approximate
upper limit of the integrated cross-section for the corresponding
sub-process (calculated in the initialization)
\begin{equation}
  \label{eq:sighatmax}
  \sigma^{\max} = \int \frac{dx_1}{x_1} \frac{dx_2}{x_2}\xf^{\max}_{1,i}(x_1) \xf^{\max}_{2,j}(x_2)
  \hat{\sigma}^{\max}_{ij\rightarrow X}(\hat{s} = S_{\max}x_1x_2),
\end{equation}
where $x_1$ is the momentum fraction of parton $i$, $\xf^{\max}_{1i}$
is an approximate upper limit of the corresponding momentum density
(and similarly for $2$ and $j$), $\hat{\sigma}^{\max}_{ij\rightarrow
  X}$ is an approximate upper limit of the hard $ij\rightarrow X$
sub-process and $S_{\max}$ is the maximum invariant mass squared of
the colliding particles.  The $\xf^{\max}$ functions are typically
normal parton density functions, but may also be convoluted with
momentum distributions of the incoming particles, if these do not have
a fixed momentum. This convolution may be in several steps, if e.g.\ 
the partons come from a resolved photon which comes from an electron
with varying momentum.

When an \hclass{XComb} has been selected, a phase-space point
$(x_1,x_2)$ is generated using the approximate parton densities and
$\sigh^{\max}$.  The chosen \hclass{XComb} is then to be accepted with
a probability given by the ratio of the exact functions to the
approximate upper limits in the phase-space point selected:
\begin{equation}
  \label{eq:weight}
  w = \frac{\xf_{1,i}(x_1)}{\xf^{\max}_{1,i}(x_1)}
  \frac{\xf_{2,j}(x_2)}{\xf^{\max}_{2,j}(x_2)}
  \frac{\sigh_{ij\rightarrow X}(S_{\max}x_1x_2)}
       {\sigh^{\max}_{ij\rightarrow X}(S_{\max}x_1x_2)}.
\end{equation}
To calculate this ratio, the \pxsec\ and \pextract\ objects may
generate internal degrees of freedom.
The Monte Carlo approximation of the integrated cross section for each
\hclass{XComb} can then be obtained by
\begin{equation}
  \label{eq:mcxsec}
  \sigma\approx\sigma_{\mbox{\scriptsize MC}} =
  \frac{N_{\mbox{\scriptsize accepted}}}{N_{\mbox{\scriptsize attempted}}}
  \sigma^{\max}.
\end{equation}
This approximation will become better the more events are generated.

At present there is no possibility to generate weighted events in
\pyth, but such facilities will be provided in the future.

\subsubsection{Constructing the event}
\label{sec:user:handler:construct}

After an \hclass{XComb} has been accepted, the rest of the generation
is assumed to continue with unit probability, although any of the
subsequent steps in the generation may veto the selected phase-space
point.

First the \pxsec\ and \pextract\ are asked to generate their internal
degrees of freedom, and the first \hclass{Step} of the event is
constructed.  This may be vetoed according to user-specified
\hclass{KinematicalCuts} which can be assigned to the \collh.

The \collh\ inherits from \partcoll, which implements the list of all
\steph s which should be applied after the initial \step\ has been set
up by the \subh. These handlers are divided into three main groups
handling parton cascades, hadronization and particle decays,
respectively\footnote{These three may, in the future, be expanded with
  a fourth group for multiple interactions.}. Each of these groups has
one special-purpose handler and two lists of \steph s to be executed
before and after the main one. The special-purpose handlers,
\cascade, \frag\ and \decayh, are all derived from the general \steph\ 
class.
  
\subsubsection{The line-mode user interface}
\label{sec:user:interface}

The \repository\ is supposed to be accessed by a user interface to
handle the setup. But the \repository\ itself also implements a
rudimentary line-mode interface which is used in the sample
\shell{setupPythia} program. The list of objects in the \repository\ 
are structured like a Unix file system with directories and
subdirectories of objects. The \repository\ communicates with these
objects via so-called interfaces which may represent a parameter, a
switch or a reference to another
object. The \textit{address} of an interface is given as\\
\hspace*{5mm}\code{/dir/subdir/.../object:interface}\\
As an example, consider an object of the class \pdata\ called
\code{Z0}, residing in a directory called \code{/Particles/}. It will
have a member variable corresponding to the nominal mass, for which an
interface is defined called \code{NominalMass}. This parameter can
then be accessed through\\
\code{/Particles/Z0:NominalMass}

The commands available to manipulate objects are listed in the
\classh{Doc/Commands.html}{Commands} file included in the program
distribution. Here is a small subset of them:
\begin{description}
  \begin{latexonly}
    \itemsep 0mm
  \end{latexonly}
\item[ls dir]~\\ print the names of the objects in the specified
  directory.
\item[describe object]~\\ print a brief description of the
  specified object, listing the names of the interfaces defined for
  it.
\item[describe object:interface]~\\ print the description for an
  interface of an object.
\item[set object:interface value]~\\ set an interface of an
  object to a given value.
\item[setdef object:interface]~\\ set an interface of an
  object to its default value.
\item[send object:command-interface message]~\\ send a message to a
  command-type (see section \ref{sec:develop:repository:Commands})
  interface of an object.
\item[get object:interface]~\\ get the value of an interface of
  an object.
\item[def object:interface]~\\ get the default value of an
  interface of an object.
\item[min object:interface]~\\ get the minimum allowed value of an
  interface of an object.
\item[max object:interface]~\\ get the maximum allowed value of an
  interface of an object.
\item[saverun run-name eventgenerator-object]~\\
  isolate an \egen\ object together with all objects it refers to,
  directly or indirectly, and save them to a file called
  \code{run-name.run}. Obviously, the changes in the repository by the
  commands above are propagated to this file.
\end{description}

\subsection{Running an event generator}
\label{sec:user:running}

When an \egen\ has been set up and saved to a file it can be used in a
number of ways.

\subsubsection{The standard run program}
\label{sec:user:running:standard}

The \shell{runPythia} program simply reads a \fullgen\ from a file and
calls it's \code{go()} method, which will run the number of events
specified in the generator. For each event a number of \anah s will be
called, again specified in the generator. In addition an \manip\ 
object may be specified e.g.\ to handle the running of several
different models of some parts of the generation on the same basic
collision.

\subsubsection{Using an event generator in another application}
\label{sec:user:running:application}

The same generator object can be read into any other application,
where each call to the \code{shoot()} method will generate an \event\ 
to be used by the application. In addition, the generator can be used
to only perform a subset of the \steph s on a user-supplied initial
\step\ using the \code{partialEvent()} method, as exemplified
in the \shell{runPartial.cc} program.

\section{Developer information}
\label{sec:develop}

In this section, some of the classes are described in somewhat more
detail to explain how one goes about implementing new models in \pyth.
Further information needed can be found in the header files of the
respective classes.

\subsection{The class structure}
\label{sec:develop:class}

The most important classes are shown in \fig{fig:handlers}. These
classes are all \emph{interfaced}, \emph{persistent} and
\emph{reference counted}. These are the base categories of classes
special to \pyth, which are used besides the standard library classes,
classes from \clhep, and small utility classes.

\subsubsection{Reference counted classes}
\label{sec:develop:refcount}

Most classes in \pyth\ are reference counted, inheriting from
\hclass{ReferenceCounted}, to avoid memory leaks in cases it is not
obvious who owns an object and is responsible for its deletion. By
using a smart pointer class \hclass{RCPtr}\footnote{defined in the
  namespace \code{Pythia7::Pointer}, which is normally imported to
  the \code{Pythia7} namespace.} together with reference counted
objects, the user need never worry about having to delete an object
when it is no longer needed - this is done automatically when there
are no more smart pointers refering to it. This means that the
syntax for creating an object is a bit modified:\\
\verb:  RCPtr<AClass> p = new_ptr(AClass());:\\
A reference counted object may still be created the normal way\\
\verb:  AClass * p = new AClass;:\\
but then someone has to be responsible for deleting the object.

Also an object which is reference counted can be pointed to by a
normal pointer:\\
\verb:  RCPtr<AClass> p = new_ptr(AClass());:\\
\verb:  AClass * p2 = p;:\\
but, of course, no guarantees are given that such a pointer will
always point to an existing object. Normal pointers could e.g.\ be
used as function arguments to avoid the overhead of incrementing and
decrementing the reference count, but it is recommended that the class
\verb:TransientRCPtr<AClass>: is used instead. This is a trivial
wrapper around a bare pointer, but can be easily assigned to/from
\verb:RCPtr<AClass>:.

There are also classes called \verb:ConstRCPtr<>: and
\verb:TransientConstRCPtr<>: for pointers to \code{const} object.
Throughout \pyth\ the pointers used are \code{typedef}'ed using a
template class defined as
\begin{verbatim}
  template <typename T>
  struct Ptr {
    typedef RCPtr<T> pointer;
    typedef ConstRCPtr<T> const_pointer;
    typedef TransientRCPtr<T> transient_pointer;
    typedef TransientConstRCPtr<T> transient_const_pointer;
  };
\end{verbatim}
i.e.\ \verb/Ptr<AClass>::pointer/ is used instead of
\verb:RCPtr<AClass>:. This is to facilitate a transition to other
smart pointer strategies.

\subsubsection{Persistent classes}
\label{sec:develop:class:persistent}

One of the serious shortcomings of the \cpp\ language is that there is
no standard way of writing objects to disk in a \emph{persistent} way
so that they can be read in again. This is especially troublesome for
a program such as \pyth, where the objects are highly interconnected.
For this reason, \pyth\ includes its own I/O system based on
\emph{persistent} streams.

A \hclass{PersistentOStream} object is able to write objects to a file
in a way such that it can be read in again by a
\hclass{PersistentIStream}. The persistent streams can read and write
basic types just as the standard \cpp\ I/O streams. But in addition
they also know how to read and write (smart) pointers to objects
derived from the \code{PersistentBase} class. They are also able to
read and write simple classes consisting of basic types and/or pointers
to \code{PersistentBase} objects if the corresponding \code{<<} and
\code{>>} operators are defined.

To make a class persistent it is not enough to inherit from
\code{PersistentBase}. One must also specify a
\hclass{ClassDescription} object, specialize the templated
\hclass{Class\-Traits} class and implement special read and write
methods. The first two items are handled semi-automatically with the
help of macros. The read and write methods must be implemented by
hand. This is, however, quite straightforward: Consider a class,
\code{AClass}, which inherits directly or indirectly from
\code{PersistentBase}. If this class has data members \code{m1},
\code{m2} and \code{m3}, which are either basic types, pointers to
persistent objects or simple classes containing either, the following
non-virtual public methods need to be defined:
\begin{verbatim}
  void AClass::persistentOutput(PersistentOStream & os) const {
    os << m1 << m2 << m3;
  }
\end{verbatim}
and
\begin{verbatim}
  void AClass::persistentInput(PersistentIStream & is, int version) {
    is >> m1 >> m2 >> m3;
  }
\end{verbatim}
where the integer argument in the latter can be used to check that the
version of the class reading in the object is the same as the one
writing it out. Note that no separating white-spaces are needed, as for
the standard \cpp\ iostreams.

The persistent streams rely on the ability to construct an object of a
class given only its name, and also to access the inheritance
relationships between classes. But since the \cpp\ standard does not
specify how this should be done -- it does not even specify a
platform-independent way of getting the name of a class -- a number of
other things need to be done for classes which are to be persistent,
as described below in section \ref{sec:develop:new}.

The \hclass{PersistentOStream} and \hclass{PersistentIStream} are
developed from the \code{Hep\-PO\-Stream} and \code{Hep\-PI\-Stream}
classes described in ref.\ \cite{CLHEP}.

\subsubsection{Interfaced classes}
\label{sec:develop:class:interfaced}

Classes which are to be handled by the \repository\ should inherit
from the \interfaced\ class, which in turn inherits from
\code{PersistentBase}. Besides the persistent read and write methods,
a concrete interfaced class must implement a \code{clone} method
returning a full copy of the object, straightforwardly implemented as:
\begin{verbatim}
  Ptr<InterfacedBase>::pointer AClass::clone() {
    return new_ptr(*this);
  }
\end{verbatim}
Optionally the class may implement protected methods called
\code{doinit} and \code{dofinish}. The \code{doinit} method is called
just before an object is used in a generator run and should be
implemented if any initialization is needed. If the initialization
fails, an \code{InitException} should be thrown. \code{dofinish} is
called for all objects used in a generator run after all events have
been generated and can e.g.\ be used to calculate and write out
statistics or, in case of \anah\ objects, to write histograms to disk.
If any of these methods are implemented they \emph{must} call the
corresponding method of the base class. If e.g.\ the initialization of
an object depends on the prior initialization of another object, the
former should call the \code{init} method of the latter. This ensures
that the \code{doinit} method is only called once per object.
Similarly, there is a non-virtual \code{finish} method for
\code{dofinish}.

\interfaced\ classes should also implement a static \code{Init} method
which will be run once for each class, in which it may create static
\code{InterfaceBase} objects to be used by the \repo\ (see section
\ref{sec:develop:repository:interfaces}).

\subsubsection{Other classes}
\label{sec:develop:class:other}

Of course, any other class allowed by the standard is allowed in
\pyth, although the implementor will manually have to handle the
deallocation of objects as well as the input and output to streams,
and these classes cannot be handled by the repository. Examples of
such classes in \pyth\ are \hclass{Lorentz5Vector} and \hclass{Selector}
(see sections \ref{sec:develop:utility:5vector} and
\ref{sec:develop:utility:select} respectively).

\subsection{The repository}
\label{sec:develop:repository}

\repository\ is a singleton class which has static lists to
keep track of and manipulate all \hclass{Interfaced} objects in the
setup-phase. There are a number of static functions defined for the
manipulation, but there is also a rudimentary command-line interfaced
as described in section \ref{sec:user:interface}. A more versatile
user interface could be implemented by inheriting from the
\repository\ class or by accessing its public static functions
from outside.

\subsubsection{Interfaces}
\label{sec:develop:repository:interfaces}

To manipulate the \hclass{Interfaced} objects in the \repository,
there are a number of interface classes defined e.g.\ to set and get
parameters in an object. These classes all inherit from the
\hclass{InterfaceBase} class. When an object of such a class is
created, it will automatically add itself to the list of interfaces in
the \repository, hence there should only be one interface object per
interface per class. This is conveniently achieved by creating static
interface objects in the static \code{Init()} method of an
\interfaced\ class.

\subsubsection{Parameters}
\label{sec:develop:repository:parameters}

To create an interface to a parameter of a class, corresponding to
e.g.\ a member variable defined as:
\begin{verbatim}
  class AClass: public Interfaced {
    // ...
  private:
    Energy energy;
  };
\end{verbatim}
the following should be done in the \code{Init()} method:
\begin{verbatim}
  void AClass::Init() {
    static Parameter<AClass,Energy>
    interfaceEnergy("Energy",
                    "This is the energy in GeV",
                    &AClass::energy, GeV,
                    1.0*GeV, 0.0*GeV, 10.0*GeV);
  }
\end{verbatim}

This will create an object interfacing the member variable
\code{energy} of the \code{AClass} class to the \repository. The
meaning of the arguments to the constructor are as follows: First the
name of the parameter used by the repository to identify it, then a
short description, followed by a pointer to the actual member, the
unit to be used when reading and writing, the default value (1~GeV),
the minimum value (0~GeV) and the maximum value (10~GeV).

There are a number of optional arguments which may be given to the
\hclass{Parameter} constructor: a flag indicating that this
parameter may only be inspected (and not changed) by the repository
(default=\code{false}), and a flag to indicate if the parameter always
should be limited to within the specified minimum and maximum values
(default=\code{true}).  Pointers to member functions for setting the
value of the parameter, and to get the current, minimum, maximum and
default values, may also be specified in the constructor, in which
case the pointer to the actual member may be the null-pointer.

There is also an interface class for vectors of parameters called
\hclass{ParVector} which can be used for interfacing any container of
parameters of a class.

\subsubsection{Switches}
\label{sec:develop:repository:switches}

To interface a switch for selecting different options in a class, the
\hclass{Switch} class can be used as follows:
\begin{verbatim}
  class AClass: public Interfaced {
    // ...
  private:
    int model;
  };

  void AClass::Init() {
    static Switch<AClass,int>
    interfaceModel("Model",
                   "Switch between different models",
                   &AClass::model, 1);
    static SwitchOption
    interfaceModelA(interfaceModel, "A", "Use model A", 0);
    static SwitchOption
    interfaceModelB(interfaceModel, "B", "Use model B", 1);
  }
\end{verbatim}
Here a \hclass{switch} object is created given a name, description,
a pointer to the actual member variable and a default value. Also a
\code{SwitchOption} object is created for each valid option, and the
arguments to the constructor is a reference to the \hclass{Switch}
object, a name, a short description and the value corresponding to the
option.

Just as for the \hclass{Parameter} class, it it possible to specify if
the switch is read-only, and it is possible to specify pointers to
member functions to set and get options rather than using the member
variable directly.

\subsubsection{Commands}
\label{sec:develop:repository:Commands}

There is a general message interface called \hclass{Command}, with
which a non-const member functions taking a \code{std::string} as
argument and returning a \code{std::string} can be made available to
the \repository:
\begin{verbatim}
  class AClass: public Interfaced {
    // ...
    string doSomething(string);
  };

  void AClass::Init() {
    static Command<AClass>
    interfaceDoSomething("DoSomething",
                         "Do something interesting",
                         &AClass::doSomething);
\end{verbatim}

\subsubsection{References between objects}
\label{sec:develop:repository:references}

If one \interfaced\ class has a pointer to an object of another
\interfaced\ class, this relation can be made available to the
\repository\ with a \hclass{Reference} object:
\begin{verbatim}
  class AClass: public Interfaced {
    // ...
  private:
    Ptr<AnotherClass>::pointer another;
  };

  void AClass::Init() {
    static Reference<AClass,AnotherClass>
    interfaceAnother("Another",
                     "A reference to another object",
                     &AClass::another);
  }
\end{verbatim}
As with \hclass{Parameter} it is possible to specify that a
\hclass{Reference} is read-only, and to specify pointers to member
functions to set and get the pointer. It is also possible to specify a
\code{rebind} flag indicating that the reference should be
rebound when exported to an \egen\ (see section
\ref{sec:develop:repository:isolation}) (default=\code{true}). It is
furthermore possible to specify a flag indicating that the pointer may
be assigned the null-pointer (default=\code{true}). Also, if the
pointer is nullable, it is possible to specify a flag --
\code{defnull} -- indicating that any null-pointer should be replaced
with a pointer to a default object when exported to an \egen\ 
(default=\code{false}).  Finally it is possible to specify a pointer
to a member function which checks if a given object will be accepted
if set.

There is also an interface for vectors of references called
\hclass{RefVector} which can be used for interfacing any container of
pointers in a class.

\subsubsection{Isolating an event generator}
\label{sec:develop:repository:isolation}

In the setup program, the \repository\ is used to connect different
objects with each other, setting their parameters and switches to
desired values. In the end the setup program should have built up a
complete \egen\ object. There may, however, exist several \egen\ 
object in the \repo, and one object may be pointed to directly or
indirectly by several \egen s. Before an \egen\ can be used it must
therefore be \emph{isolated} from the other objects in the repository.

The isolation procedure starts by cloning the specified \egen\ and all
(\interfaced) objects which are referenced directly or indirectly by
the \egen. To find out which objects are needed, the \repo\ checks all
\hclass{Reference} and \hclass{RefVector} interfaces of the \egen. The
objects found in that way are again checked for \hclass{Reference} and
\hclass{RefVector} interfaces, and so on. If an \interfaced\ class has
non-interfaced pointers to other objects, these pointers should be
communicated to the \repo\ using the virtual \code{getReferences()}
method (declared in \hclass{Interfaced\-Base}, the base class of
\hclass{Interfaced}). Note that if the \code{getReferences()} method
is implemented for a class, this implementation must call the
\code{getReferences()} of the base classes.

Cloning the objects is not enough, however, since a cloned object may
still be pointing to objects in the repository rather than to their
clones. Therefore all pointers must be \emph{rebound} to point to the
correct clones. For pointers which are interfaced with
\hclass{Reference} or \hclass{RefVector} (which do not have the
\code{norebind} flag set), this rebinding is automatic. For other
pointers this must be done via the\\
\verb:virtual void rebind(const TranslationMap & trans):\\
function (declared in \hclass{InterfacedBase}). The argument to this
function is a \hclass{Rebinder} (via a typedef) and should be used
like this:
\begin{verbatim}
  class AClass: public ABaseClass {
    // ...
  private:
    Ptr<AnotherClass>::pointer another;

  protected:
    virtual void rebind(const TranslationMap & trans)
        throw(RebindException) {
      another = trans[another];
      ABaseClass::rebind(trans);
    }
  };
\end{verbatim}
Note that if this function is implemented it must call the
\code{rebind} method of the base class.

Finally the \egen\ goes through all reference interfaces of all cloned
objects and if one is found which is set to the null-pointer, and if
the \texttt{defnull} flag is set, and an object of the correct type is
available in the \egen s list of default objects, that object will be
assigned to the interface.

After an \egen\ has been isolated, all the objects that were cloned
and included in the run will have a pointer to this \egen. Though this
pointer an \interfaced\ object will have access to global properties of
the run. It is important to note the difference between the setup
phase and the run phase in this respect. If, during the setup phase,
an object wants to access a \pdata\ object it should do so using
static methods of the \repo, while during the run phase it should use
its \egen\ pointer and use the corresponding non-static members of the
\egen.

\subsection{Creating new handler classes}
\label{sec:develop:new}

When implementing a physics model in \pyth, the procedure is to create
a new class inheriting from one of the existing \emph{handler}
classes, overriding the relevant virtual functions. Since all handler
classes inherits from \interfaced\ which in turn is persistent, there
are a number of common things to do for any new handler class
\code{AClass}:
\begin{itemize}
  \itemsep 0mm
\item The class must have a public default constructor.
\item Exactly one object of either class
  \verb:ClassDescription<AClass>:,\\
  \verb:AbstractClassDescription<AClass>:,
  \verb:NoPIOClassDescription<AClass>:\\
  or \verb:AbstractNoPIOClassDescription<AClass>: must be instantiated
  to register this class with the persistent stream classes.  This is
  typically achieved by a static member. Which of the description
  classes to choose depends on whether \code{AClass} has members which
  must be written persistently and whether it is abstract or concrete.
\item The templated \hclass{ClassTraits} should be specialized to
  supply information about this class to the persistent streams.
  At least the static\\
  \verb/ClassTraits<AClass>::className()/ method must be overridden to
  return a unique and platform-independent string representing the
  name of \code{AClass}. This is conveniently done using the
  \code{PYTHIA7\_DECLARE\_CLASS\_TRAITS} macro.
\item The templated \code{BaseClassTrait} must be specialized for each
  immediate base class which is persistent so that
  \verb/BaseClassTrait<AClass,1>::NthBase/ is a typedef for the first
  base class, \verb/BaseClassTrait<AClass,2>::NthBase/ a typedef for
  the second, and so on. This is needed by the persistent streams
  and is conveniently done using the
  \code{PYTHIA7\_DECLARE\_CLASS\_TRAITS} macro.
\item If the class has members which need to be written and read
  persistently, the (non-virtual) methods\\
  \verb:void persistentOutput(PersistentOStream &) const: and\\
  \verb:void persistentInput(PersistentIStream &, int):\\
  must be implemented. (The names of these functions can be controlled
  in the \hclass{ClassTraits} class.)
\item If the class is concrete it must implement a clone method as
  described in section \ref{sec:develop:class:interfaced}.
\item If the object produced from a \code{clone()} call needs some
  initialization to be useful in the \repo, the \code{fullclone()}
  should be implemented to return an initialized object.
\item If the objects of this class needs to be initialized before an
  \egen\ run is started, the
  \verb:void doinit() throw (InitException): should be implemented
  (which must always must call the \code{doinit()} method of the
  base classes). If the initialization fails, the method should throw
  an \code{Init\-Excep\-tion}.
\item If the objects of this class needs to do something after an
  \egen\ run, e.g.\ write out statistics, the \verb:void dofinish():
  method should be implemented (which must call the
  \code{dofinish()} method of the base classes).
\item The class must implement a static \code{Init()} method, possibly
  containing static \hclass{Inter\-face\-Base} objects as described in
  section \ref{sec:develop:repository}. This method is called once for
  each class in the construction of the static \code{ClassDescription}
  object of that class.
\item If the class has pointers to other objects which are not
  interfaced, the method
  \verb|vector<Ptr<InterfacedBase>::pointer> getReferences()|
  must be implemented. The returned vector should
  contain the result from the call to \code{getReferences()} of the
  base class, plus all non-interfaced pointers in \code{AClass}.
\item If the class has pointers to other objects which are not
  interfaced, the method
  \verb:void rebind(const TranslationMap & trans):
  must be implemented, calling the \code{rebind()} method of
  the base class using the argument as described in section
  \ref{sec:develop:repository:isolation}.
\end{itemize}
This may seem like a lengthy and cumbersome procedure just to create
one new class. But typically only a few of these items are necessary,
and most things can be generated automatically. \pyth\ contains a
number emacs-lisp functions (in the \shell{Templates} directory) for
easy generation of complete skeletons for new classes, containing
default versions of the methods needed.

\subsubsection{Partonic cross-sections}
\label{sec:develop:new:xsec}

The \pxsec\ class is the handler class from which all
classes implementing hard $2\rightarrow n$ partonic cross sections
should be inherited. In the basic form a \hclass{PartonXSecFn} should
represent the function $\hat{\sigma}_{ij\rightarrow X}(\hat{s})$ so
that the total integrated cross section for the corresponding
sub-process is given by:
\begin{equation}
  \label{eq:sighat}
  \sigma = \int \frac{dx_1}{x_1} \frac{dx_2}{x_2} \xf_{1,i}(x_1) \xf_{2,j}(x_2) \hat{\sigma}_{ij\rightarrow X}(S_{\max}x_1x_2),
\end{equation}
where $x_1$ is the momentum fraction of parton $i$ and $\xf_{1i}$ is
the corresponding momentum density (and similarly for $2$ and $j$).

But \hclass{PartonXSecFn} is more than a simple function and to be
usable in the sub-process selection described in section
\ref{sec:user:handler:subp}, there are a number of different virtual
member functions which can be specialized.
\begin{itemize}
  \begin{latexonly}
    \itemsep 0mm
  \end{latexonly}
\item{\bf\verb|sigHatMax(const SInterval & S, const SInterval & SHat,|\\
  \verb|     const cPDPair &, const KinematicalCuts &) const;|}\\
  should return vector of function objects of class
  \hclass{Sigma\-Hat\-Max\-Base} corresponding to $\sigh^{\max}$ in eq.\ 
  \ref{eq:sighatmax}. Each of these objects should correspond to a
  term with a specific momentum and colour geometry (see below). The
  arguments are an interval in $S$, an interval in $\shat$, the pair
  of incoming partons and a \hclass{KinematicalCuts} object. This
  method is called in the initialization phase and it should be noted
  that it is a \code{const} function, since the same object may be be
  used for several different sub-processes. Any process-dependent
  initialization should therefore be stored in the returned
  \hclass{Sigma\-Hat\-Max\-Base} objects.
\item{\bf\verb|pdfScale();|}\\
  is called after a phase-space point has been generated by the
  \collh. It should return the scale of the hard sub-process to be
  used in the parton density functions. No arguments are given, but
  since \pxsec\ inherits from the \lastinfo\ class, all information
  about the selected phase-space point is already available. The
  \pxsec\ typically needs to generate internal degrees of freedom to
  be able to calculate the scale, and these should then be saved for
  subsequent calls.
\item{\bf\verb|weight();|}\\
  should return $\sigh(\shat)/\sigh^{\max}(\shat)$, the ratio between
  the exact cross section function and the approximate upper limit at
  the generated phase-space point.  This ratio should be less than
  unity. But occasional weights above one can be handled by a
  compensation mechanism in the \collh. If the approximate cross
  section is the result of an integral over internal degrees of
  freedom, $\sigh^{\max}=\int dz_1\cdots dz_n
  d\sigh^{\max}(\shat,z_1,\ldots,z_n)/dz_1\cdots dz_n$, this function
  may generate some or all of these degrees of freedom and return the
  weight
  \begin{equation}
    w=\frac{\frac{d\sigh(\shat,z_1,\ldots,z_n)}{dz_1\cdots dz_n}}
    {\frac{d\sigh^{\max}(\shat,z_1,\ldots,z_n)}{dz_1\cdots dz_n}}.
    \label{eq:sigweight}
  \end{equation}
  Again, no arguments are needed. If internal degrees of freedom are
  generated, these should be saved for future use.
\item{\bf\verb|construct(tSubProPtr sub);|}\\
  is called if the selected phase-space point has been accepted by the
  \collh, and takes a pointer to a \subprocess\ as argument. In this
  function the complete hard parton--parton sub-process should be
  constructed using the internal degrees of freedom previously
  generated in \code{pdfScale} and/or \code{weight} together with the
  remaining ones that need to be generated here.
\end{itemize}

The momentum geometry for a general $2\rightarrow n$ process is
described in terms of a vector of $n-1$ sets of four integers, each
set corresponding to one of the internal lines (a four-parton vertex
is treated as an internal line in this respect). The first two numbers
corresponds to the indices of the partons of one of the vertices, the
last two to the other. The incoming partons are numbered 1 and 2, the
outgoing $3\ldots n+2$, and the internal lines $n+3\ldots 2n+1$.  This
geometry is in most cases completely unphysical -- it is not possible
to differentiate between different Feynmann diagrams with the same
final state -- and can be left out. However, especially if a
$2\rightarrow n$ sub-process is followed by a partonic cascade,
information about which geometry is the most likely may be necessary.

The colour geometry is given as a vector of $2+n$ numbers
corresponding to the flow of colour from each of the incoming and
outgoing partons. Also so-called colour-junctions connecting three
colour triplets (or three anti-colour triplets) to each other, can be
described by specifying (anti-) colour flow to an imaginary parton
with index $>2+n$.

\pxsec\ is the most general form for a hard $2\rightarrow n$ partonic
cross-section. Simple $2\rightarrow 1$ sub-processes are easy to
implement using this base class, although the handling of resonances
with the \textit{width generator} mentioned in section
\ref{sec:user:particle} has not yet been implemented properly. For
$2\rightarrow 2$ processes there is a special base class called
\hclass{THatXSecFn} inheriting from \pxsec\ and representing
$d\sigh(\shat,\that)/d\that$.

The virtual functions which needs to be implemented in
\hclass{THatXSecFn} are as follows.
\begin{itemize}
  \begin{latexonly}
    \itemsep 0mm
  \end{latexonly}
\item{\bf\verb|sigHatMax(const SInterval & S, const SInterval & SHat,|\\
  \verb|     const cPDPair &, const KinematicalCuts &) const;|}\\
  should now return \hclass{THatFunctionBase} objects which inherits
  from the \hclass{Sigma\-Hat\-Max\-Base}. This class automatically takes
  care of the integral over and generation of $\that$ given any
  $d\sigh(\shat,\that)/d\that$ function (although special integration
  and generation of course may be implemented in derived classes).
\item{\bf\verb|getOutgoing() const;|}\\
  should return the types of the pair of outgoing partons for the
  selected phase-space point and colour and momentum geometry.
\item{\bf\verb|getIntermediate() const;|}\\
  should return the exchanged parton for the selected pair of outgoing
  partons, the selected phase-space point and colour and momentum
  geometry.
\item{\bf\verb|reweight();|}\\
  is called by the \code{weight} function to give any weight related
  to approximations made when giving the \hclass{THatFunctionBase}
  objects in the \code{sigHatMax} method. One example is to return the
  ratio between the $\alpha_{\mbox{\scriptsize S}}$ at the scale of
  the chosen phase-space point and the maximum value which may have
  been used in the initialization.
\end{itemize}
A class inheriting from \hclass{THatXSecFn} may, of course, also
override other virtual functions of \pxsec\ if needed. 

A general base class for $2\rightarrow n$ processes, where only the
squared matrix elements for different colour and momentum geometries
need to be implemented will be provided in the future.

\subsubsection{PDF's and remnant handling}
\label{sec:develop:new:pdfs}

All parton density functions should be implemented in classes
inheriting from \pdfbase. Just as for \pxsec, the \pdfbase\ class
should implement both an approximate upper limits of the density
functions for each separate parton in a particle, as well as the
functions themselves. The approximate upper limits shall be given as
\hclass{ApproxPDF} objects, each of which represents a sum of powers
of $\log(1/x)$, the logarithmic momentum fraction:
\begin{equation}
  \label{eq:approxpdf}
  \xf^{\max}=\sum_i c_i \log(\frac{1}{x})^{a_i}.
\end{equation}
For a given $x$, this function should be larger than the true function
for any scale allowed by the kinematical cuts. The \pdfbase\ class has
methods to automatically produce such approximations, but derived
classes may specify their own.

Also the momentum distribution of the incoming
particles should be approximated by \hclass{ApproxPDF}s, making it
simple to convolute with these. Even in complicated cases such as
distribution of partons in a pomeron in a photon in an electron with
varying momentum, it is easy to convolute \hclass{ApproxPDF}s to get
the total parton momentum densities:
\begin{equation}
  \label{eq:folding}
  \xf^{\max}_{\mbox{\scriptsize tot}}(x)=\int_x^1 \frac{dx'}{x'}
  x'\!f^{\max}_1(x')\frac{x}{x'}\!f^{\max}_2(\frac{x}{x'}) =
  \sum_{i,j} c_{1i}\cdot c_{2j} \log(\frac{1}{x})^{a_{1i}+a_{2j}+1}.
\end{equation}

The important virtual methods of \pdfbase\ are as follows:
\begin{itemize}
  \begin{latexonly}
    \itemsep 0mm
  \end{latexonly}
\item{\bf\verb|canHandleParticle(tcPDPtr particle) const;|}\\
  should return \code{true} only if the implemented PDF can be used
  for the particle type given in the argument.
\item{\bf\verb|partons(tcPDPtr particle) const;|}\\
  should return a vector of partons which may be extracted from the
  particle type given in the argument.
\item{\bf\verb|approx(tcPDPtr particle, const PDFCuts &) const;|}\\
  given a particle and an object representing the ranges in momentum
  fraction and scale where this PDF will be used, this method should
  return a map of \hclass{ApproxPDF} objects indexed by the
  corresponding parton type.
\item{\bf\verb|xfx(tcPDPtr particle, tcPDPtr parton, Energy2 scale,|\\
  \verb|double x, double eps = 0.0, Energy2 offshell = 0.0*GeV2) const;|}\\
  should return the momentum density, given as arguments the particle,
  the parton, the scale at which the parton is resolved, the momentum
  fraction $x$ of the parton and, optionally, \code{eps}$=1-x$ (for
  precision reasons when $x$ is close to 1) and the offshellness of
  the particle (mainly intended for resolved virtual photons).
\item{\bf\verb|xfvx(tcPDPtr particle, tcPDPtr parton, Energy2 scale,|\\
  \verb|double x, double eps = 0.0, Energy2 offshell = 0.0*GeV2) const;|}\\
  may be implemented in the same way as \code{xfx} to give only the
  valence part of the momentum density. The default version simply
  returns 0.
\end{itemize}

\pdfbase\ has a pointer to a \remh, which should be able to construct
the remnants for any parton which may be extracted from a given
particle. \remh\ has the following virtual functions:
\begin{itemize}
  \begin{latexonly}
    \itemsep 0mm
  \end{latexonly}
\item{\bf\verb|canHandle(tcPDPtr particle, const cPDVector & partons) const;|}\\
  should return \code{true} if the \remh\ can handle the remnant
  generation for a given particle type and a vector of partons to be
  extracted.
\item{\bf\verb|getRemnants(tcPDPtr particle, tcPDPtr parton, double x,|\\
    \verb|       Energy2 sMax, TransverseMomentum & kt)|}\\  
  should return a vector of remnants. The arguments are the particle,
  the extracted parton, the momentum fraction and an upper limit on
  the invariant mass squared of the remnants. The remnants should be
  returned in their own center-of-mass system assuming that the
  incoming particle is along the positive $z$-axis.  The intrinsic
  transverse momentum of the extracted parton may also be returned
  through a reference argument.
\end{itemize}
A second version of the \code{getRemnants} will be supplied in the
future, implementing multiple parton extractions from a particle.

The whole process of extracting partons from particles and creating
remnants is administered by a \pextract\ object, which has two main
virtual functions:
\begin{itemize}
  \begin{latexonly}
    \itemsep 0mm
  \end{latexonly}
\item{\bf\verb|weight();|}\\
  should return the product of the ratios of the true to the
  approximate parton density functions used for the selected
  phase-space point. The function takes no argument and all
  information about the selected phase-space point can be obtained
  from the \lastinfo\ base class.
\item{\bf\verb|construct(tCollPtr, tStepPtr, tSubProPtr)|}\\
  given pointers to the current \collision, the first \step\ of that
  collision and the \subprocess, the complete kinematics of the hard
  sub-process should be constructed and be put into the \step.
\end{itemize}
The \pextract\ also has a number of methods to be used by subsequent
step handlers to access information about the parton densities and
remnant handlers.

\subsubsection{Luminosity functions}
\label{sec:develop:new:lumifn}

The \lumifn\ base class should be used to describe the momentum
distribution of the beam particles. As for the parton densities, these
should be given both as an approximate upper limit in the form of
\hclass{ApproxPDF} objects and as a member function giving the exact
distribution. The following virtual methods should be implemented by a
derived class:
\begin{itemize}
  \begin{latexonly}
    \itemsep 0mm
  \end{latexonly}
\item{\bf\verb|canHandle(const cPDPair &) const;|}\\
  should return true if the class can handle the pair of incoming
  particles given as arguments.
\item{\bf\verb|getSBins(const cPDPair &) const;|}\\
  should return a vector of intervals in total particle--particle
  invariant mass squared, $S$, for which this class can be used.
\item{\bf\verb|probDists(const SInterval &) const;|}\\
  should return a pair of \hclass{ApproxPDF}s giving the approximate
  upper limits of the momentum distributions of the incoming particles
  for the interval in $S$ given as argument.
\item{\bf\verb|weight();|}\\
  for the selected phase-space point (which is available through the
  \lastinfo\ base class) return the product of the ratios of the true
  to the approximate momentum distribution functions.
\end{itemize}

\subsubsection{Step handlers}
\label{sec:develop:new:step}

After the initial step of a \collision\ has been generated, a sequence
of \steph s are called to perform steps necessary to complete the
generation of one collision. The \steph\ base class is used for any
kind of such process and has only one virtual method to be overridden
by derived classes:
\begin{itemize}
  \begin{latexonly}
    \itemsep 0mm
  \end{latexonly}
\item{\bf\verb|handle(PartialCollisionHandler & ch, const tcPVector & tagged,|\\
    \verb|  const Hint & hint);|}\\
  is given a reference to the current \partcoll, a vector of
  \emph{tagged} particles and a \hint. The idea is that the \steph\ 
  should examine the tagged particles and the hint to see if there is
  anything that it can do. If so, it should do whatever it is supposed
  to do and, if this results in new particles being produced or old
  ones being modified, it should get a copy of the last \step\ by
  calling the \code{newStep()} method of the \partcoll, and put the
  resulting particles there.
\end{itemize}
Although only the tagged particles are directly available in the
\code{handle} method, the \steph\ is free to manipulate any part of
the event produced so far which is available from the \partcoll\ using
the \code{currentEvent()} method.

The basic \hint\ class simply contains a list of tagged particles and
a scale. The list of particles in the \hint\ is almost the same as the
ones passed as argument to \code{handle}, but the latter only
contains particles which are still present in the last \step, when
\code{handle} is called. The meaning of the scale in the \hint\ is
unspecified -- the implementor of a \code{handle} method should
clearly document how this scale is interpreted. A \steph\ class can
have tailor-made \hint\ classes, in which case the \code{handle}
method should \code{dynamic\_cast} the \hint\ supplied in the
argument. Note, however, that this cast must be checked manually since
there is no guarantee that the hint is of the expected class.

A special \hint\ object is the \emph{default} hint, obtainable from
the static \code{Default} method in the \hint\ class, and
corresponds to telling the \steph\ \textit{Look through the current
  event, check if there is anything to do and do it.}

Note that a \steph\ does not need to do anything. In fact, also things
like analysis and different kinds of cuts can be implemented as a
\steph. In the latter case, the \steph\ can look if the event would
fail some cuts, in which case it should throw a \code{Veto}
exception - causing the \eventh\ to throw away the current event and
generate a new one. Alternatively, the \steph\ may throw a
\code{Stop} exception, in which case the \eventh\ will stop
generating and return the event generated so far to the calling
program.

Any one can at any time during the generation of an event add a
\steph\ and a \hint\ to a \partcoll\ using the \code{addStep}
method. Here one should also specify to which group of handlers the
\steph\ should be added and at which level in that group. The group
can be given as \code{Group::subpro},
\code{Group::cascade}, \code{Group::hadron} or
\code{Group::decay} and the level can be given as
\code{Group::before}, \code{Group::main} or \code{Group::after}.
Specifying e.g.\ \code{Group::cascade} and \code{Group::before}
(\code{Group::after}) will add a \steph\ and a \hint\ to the list of
handlers to be called before (after) calling the \cascade, while
specifying \code{Group::cascade} and \code{Group::main} will
replace the current \cascade\ (if not the null-pointer is specified)
and add the \hint\ to the list of hints for which the main \cascade\ 
will be called. Note that if the cascading has already been performed,
the \partcoll\ will restart the cascading in the next step.

When the generation of an event is started, the groups of step
handlers will be filled with handlers which are given \emph{default}
hints. These handlers are specified in the setup phase in the
\partcoll. It is also possible to specify handlers in the different
\subh s. The default handlers specified in the \subh\ selected for an
event will take precedence over the ones in the \partcoll.

The groups are then processed in order, first the list of
post-sub-process handlers\footnote{Note that there are no
  pre-sub-process handlers and that the main sub-process is handled
  outside of this structure}, then the list of pre-cascade handlers,
followed by the main \cascade\ possibly called several times with
different hints, then the post-cascade handlers, continuing in the
same way with the hadronization and decay groups.

\subsubsection{Cascade handlers}
\label{sec:develop:new:cascade}

So far no model for QCD cascades has been implemented in \pyth, and
the \cascade\ class currently does not introduce any functionality
beyond the \steph\ base class.

\subsubsection{Hadronization handlers}
\label{sec:develop:new:hadronization}

The Lund string fragmentation model was the first substantial physics
module to be added to \pyth. The implementation is discussed briefly
in section \ref{sec:lundfrag} and in detail in ref.\ \cite{Marc}.

\subsubsection{Decay handlers}
\label{sec:develop:new:decay}

The \decayh\ should be used to administer the decay of unstable
particles. The actual generation of phase space for decay products is
handled by the \decayer\ assigned to each decay channel. The \decayh\ 
base class implements the \code{handle} method declared in \steph,
where it goes through the tagged particles and, for the unstable ones,
selects a decay channel, asks the \decayer\ to perform the actual decay
and inserts the children in the new \step, setting up
mother--daughter relationships, etc. Children which are unstable are
again decayed, until only stable particles are left.

Note that the decay of an unstable particle may also be done inside
other step handlers using the \decayer\ objects given in the decay
table of a \pdata\ object. This is useful e.g.\ in a \cascade\ where
the decay of a top quark may be required in the middle of the cascade.

\subsubsection{The Standard Model}
\label{sec:develop:new:sm}

The \egen\ has a pointer to a \sm\ object which carries information
about the Standard Model parameters to be used in a run. \sm\ has
\code{inline}d functions to access information which can be calculated
at initialization time. To calculate some parameters, \sm\ has
pointers to objects implementing $\alpha_{\mbox{\scriptsize EM}}$,
$\alpha_{\mbox{\scriptsize S}}$ and the CKM matrix using the
\hclass{Alpha\-EM\-Base}, \hclass{Alpha\-S\-Base} and
\hclass{CKM\-Base} classes respectively.

So far, there is no \emph{beyond} Standard Model physics in \pyth. But
it is natural to implement e.g.\ SUSY parameters in classes inheriting
from \sm. Handler classes which need access to SUSY parameters, would
then dynamically cast the \sm\ object available through the \egen\ to
a SUSY parameter object to access these parameters.

\subsubsection{Random Numbers}
\label{sec:develop:new:rnd}

The \egen\ has a pointer to a \rnd\ object which can be used by any
\interfaced\ object through the inlined \code{rnd()} method
returning a random number in the range $]0,1[$. Also methods for
getting random numbers in other ranges, as well as random integers are
given.

\rnd\ only defines the interface to a random generator and assumes
that derived classes will handle classes derived from the
\code{RandomEngine} class of \clhep. A derived class need only
implement the virtual \code{randomGenerator} method returning a
reference to the \code{RandomEngine} object.

Generating random numbers is a very central part in event generation
and typically 10-20\% of the total running time is spent generating
random numbers. To reduce the overhead of having to call virtual
functions every time a random number is needed, the \rnd\ uses the
\code{flatArray} method of the \code{RandomEngine} object to
generate a large number of numbers which are cached in a large vector,
In this way an inlined member function of \rnd\ can be used to get a
random number, and only once every thousand calls or so, a virtual
method needs to be called to fill the cache.

\subsubsection{Exceptions}
\label{sec:develop:new:ex}

The \hclass{Exception} class should be used for all exceptions thrown
by a handler which are supposed to be caught by either the \repo\ in
the setup phase or by the \egen\ in the running phase. A class
deriving from \hclass{Exception} should, in the constructor, specify
the kind of error which has occurred using the \code{severity()}
member function and write a message to the protected
\code{theMessage} member variable of type
\code{std::ostringstream}.

The error types are enumerated as follows:
\begin{description}
  \begin{latexonly}
    \itemsep 0mm
  \end{latexonly}
\item[unknown]~\\ if no type was specified, the program will
  abort and dump core as soon as the exception is thrown.
\item[info]~\\ this is not really an error and the exception should in
  principle not be thrown. Instead it should be logged using the
  \code{logException} method of the \egen\ class which simply writes
  out the message to a log file. It is possible to fix the maximum
  number of messages of the same kind which will be written out.
\item[warning]~\\ should be treated the same way as the
  \code{info} exception, i.e.\ it should be logged with the \egen.
\item[eventerror]~\\ the generation of the current event will be
  aborted. The exception will be caught by the \egen\ which may decide
  to abort the execution if too many exceptions of a given type is
  caught.
\item[runerror]~\\ the whole run will be aborted but the
  exception is caught by the \egen\ who gracefully returns the control
  of the execution to the calling process.
\item[maybeabort]~\\ a serious error has been found which may
  depend on a programming error. \egen\ catches this exception to
  write out a message but then re-throws the error so that the calling
  function may catch it and continue executing.
\item[abortnow]~\\ the message is written directly to
  \code{std::cerr} and the process is aborted and a core dump is
  produced before the throwing procedure has started the stack
  unwinding.
\end{description}

In the end of each run, a summary of all exceptions which have
occurred is given in the log-file (class name and count).

\subsection{Utility classes}
\label{sec:develop:utility}

\pyth\ contains a number of small utility classes, and surely the
number will increase as more physics models are implemented in the
framework. Some of these classes are not specific to \pyth, and they
should maybe migrate into \clhep, others are more specific. Here are
some examples:

\subsubsection{Lorentz5Vector}
\label{sec:develop:utility:5vector}

\lvec\ inherits from the \code{LorentzVector} class of \clhep, and
simply adds a data member implementing the mass component. The reason
is not just to cache the mass component for easy access, it is also
useful to avoid precision problems \footnote{even if double precision
  is used, a TeV neutrino may e.g.\ have unacceptably large error in
  its invariant mass} and in general it can be technically convenient
to have a mass component different from the invariant mass.

\subsubsection{Math functions}
\label{sec:develop:utility:math}

Inside the namespace \classh{Pythia7::Math}{Math} there are a number of
mathematical functions and classes defined. Here are some examples:
\begin{itemize}
  \begin{latexonly}
    \itemsep 0mm
  \end{latexonly}
\item{\bf\verb|gamma(double), lngamma(double)|}\\
  returns the (log of the) gamma function.
\item{\bf\verb|exp1m(double)|}\\
  returns $1-\exp(x)$ to highest possible precision for $x$ close to
  0.
\item{\bf\verb|log1m(double)|}\\
  returns $\log(1-x)$ to highest possible precision for $x$ close to
  0.
\item{\bf\verb|template <int N> double Pow(double)|}\\
  returns the argument raised to the integer power given as the
  template argument.
\end{itemize}

\subsubsection{Selector}
\label{sec:develop:utility:select}

The \hclass{Selector} class declared as
\begin{verbatim}
  template <typename T, typename WeightType = double>
  class Selector;
\end{verbatim}
stores objects of type \code{T} and associates them with a weight.
Using the \code{select} method, giving random numbers in the range
$]0,1[$, objects are retrieved one at the time with a probability
proportional to their weight.

\subsubsection{SimplePhaseSpace}
\label{sec:develop:utility:phase}

The \hclass{Simple\-Phase\-Space} class has a number of static member
functions to distribute two or three particles in phase space in their
center of mass system. The methods are called \code{CMS} and take
two or three references to particles and a total invariant mass
squared as argument, together with angles and, in the case of three
particles, energy fractions. In some cases the angles may be generated
isotropically given an \rnd\ as argument.

\subsubsection{Other utility classes}
\label{sec:develop:utility:other}

\begin{itemize}
\item \hclass{Interval} is a templated class representing an interval
  of numbers corresponding to the template argument type. E.g.\\
  \verb|Interval<double> interval(2.2, 13.0)| represents the interval
  $[2.2,13.0[$
\item \hclass{HoldFlag} is a templated class used to give temporary
  values to a variable in an exception-safe way such that the old
  value is recovered when the \hclass{HoldFlag} object is destroyed.
\item \hclass{Triplet} is a templated class completely analogous to
  the standard \code{std::pair} class.
\end{itemize}

\subsection{Documentation}
\label{sec:develop:documentation}

Besides this document, the main documentation of \pyth\ is found in
the header files. These are commented in a way such that they can be
automatically converted into HTML pages using a small \emph{awk} script
included in the distribution.

For the handler classes there is also some documentation provided in
the definition of the interfaced parameters and switches. In the
future it should be possible to extract this information to produce a
brief description of an \interfaced\ class and its interfaces in HTML
and \LaTeX\ format.

For handler classes implementing a specific physics model, it is
possible to communicate information about this model using the
virtual \code{modelDescription} and \code{modelReferences}
functions defined in the \interfaced\ class.
\begin{itemize}
  \begin{latexonly}
    \itemsep 0mm
  \end{latexonly}
\item{\bf\verb|modelDescription();|}\\
  should return an \code{std::string} containing a \LaTeX\ 
  \verb|\item| with a brief description of the model.
\item{\bf\verb|modelReferences();|}\\
  If \code{modelDescription} returns a text with citations, the
  corresponding\\ \verb:\bibitem:s should be returned here.
\end{itemize}
If a class implements any of these two functions the corresponding
methods in the base class should also be called and the result
concatenated. After an \egen\ run, the \code{modelDescription} and
\code{modelReferences} methods of all objects which has been
used\footnote{To indicate that an object has actually been used, the
  \code{useMe()} method defined in the \interfaced\ class should be
  called at least once.} will be called and the result will be put in
a \code{.tex} file from which a user can cut and paste the relevant
references into a publication where results from \pyth\ has been used.

\section{Case study : The Lund String Fragmentation}
\label{sec:lundfrag}

In this section, we briefly describe the implementation of the Lund
fragmentation model that will be the main option for hadronization in
\pyth. We focus on the design analysis and present the key components
of its implementation. A complete description of the analysis and
design is given in \cite{Marc}.

\subsection{The Lund fragmentation algorithm }
\label{sec:lundfrag:algorithm}

In the Lund model, the colour field between two colour-connected partons 
is approximated by a massless relativistic string, which can break  
by the creation of $q\bar{q}$ pairs to produce hadrons 
(see \cite{And83, And98} for a complete review on the Lund model). 

In the case of generic multiparton states, such as those produced
e.g.\ by parton showering in $e^+e^-$ annihilation, the string is
stretched from the quark to the antiquark endpoints via the
colour-connected gluons, which can be considered to be internal
excitations of the string field. The mechanism describing the
fragmentation of such a system is very complicated and presented in
detail in \cite{sjo84, sjo94}. In this section, we briefly introduce
the main features of the Lund algorithm that will lead to the class
category identification.

For a Monte Carlo implementation, the fragmentation algorithm is
conveniently expressed in the momentum space representation, where it
can be formulated as an iterative sequence of steps starting from both
{\it endpoints} and proceeding towards the middle of the string
\cite{sjo84}. Each step $(i)$ is taken from the {\it last} endpoint,
previously produced in the same direction, to a new production vertex
point where a new $q_i\bar{q_i}$ pair is created. The flavour
$\bar{q}_{i-1}$ left at the last endpoint then pair off with the new
$q_{i}$ to form the new hadron. The left-over flavour $\bar{q}_{i}$
then form the new starting endpoint for the next step. These iterations
proceed until the remaining mass of the string is below a minimum
value, at which point two final hadrons are produced.

\subsection{The design analysis}
\label{sec:lundfrag:design}

The current design of the Lund fragmentation modules is structured
around two main class categories :

\begin{description} 
\item[{\it The string representation} : ] Encapsulates the string
  information and provides a flexible framework for the fragmentation
  algorithm to operate on.  The main strategy used for the design
  development was to decouple the category responsible for the string
  fragmentation administration from any particular string
  representation. Classes using a string object should not have
  explicit knowledge of its internal structure but rather access
  information through well-defined interfaces.
\item[{\it The fragmentation administration} : ] To mirror the
  analysis of section \ref{sec:lundfrag:algorithm}, the fragmentation
  management should results in a step-by-step updating procedure of
  endpoints, each step corresponding to a new hadron creation.  This
  is achieved by the development of an \endp\ type used by the
  fragmentation administrator (the \lundfragh) to manage the sequence
  of steps.  To allow for improved code maintainability and
  flexibility in terms of new physics developments, the \lundfragh\ 
  should not depend on the \endp\ implementation and should use it as
  an ``iterator-like'' type in the management of the iterative
  procedure.
\end{description} 

\subsection{The \lundfragh\  class}
\label{sec:lundfrag:fragh}

The \lundfragh\ class is responsible for the administration of the
fragmentation procedure. The main components are :
\begin{itemize}
\item A \String\ object that holds all information about the current
  string to hadronize.
  
\item Three \endp s : to describe the last two endpoints previously
  produced from the ends of the string, and the new endpoint of the
  current step being performed.  Each \endp\ encapsulates information
  about the endpoint parton type and the breakup vertex position.
  
\item Pointers to the \lundptgen, \lundzgen, \lundflavourgen\ objects
  for the $p_t, z$ and flavour generation.

\item A list of \hadron s holding information on the newly created hadrons. \\
  In the process of fragmenting the \String, the \lundfragh\ will
  often try several times to complete the hadronization.  To avoid the
  extra overhead of creating and destroying \particle\ objects,
  \hadron\ provides a simplified implementation which contains the
  type and momentum of a temporary particle.
\end{itemize}

\noindent 

To make the \lundfragh\ class reusable, we split the string-breaking
administration into separate methods.

\begin{itemize}
\item{\bf\verb|handle(PartialCollisionHandler & ch, const tcPVector & tagged,|\\
    \verb|  const Hint & hint);|}\\
  This is the only public method of the \lundfragh\ and it overrides
  the one inherited from the \frag\ base class (see sections
  \ref{sec:develop:new:step} and \ref{sec:develop:new:hadronization}).
  Its task is to look through the list of {\it tagged} particles in
  the current \step, extract the strings if any, and send them to the
  \code{hadronize} method to be fragmented. At the end of the
  fragmentation, it gets from the \collh\ a copy of the last \step, in
  order to add to the event record the newly created particles before
  informing the \collh\ to continue the generation. This procedure is
  very general and will probably be moved to the \code{handle} method
  of the \frag\ class.
\item{\bf\verb|Hadronize(const tPVector &);|}\\
  Is the major method in the string fragmentation administration.
  Given a vector of particles, received from the \code{handle} method,
  it returns the list of created hadrons. Its administration task can
  be summarized as follows :
  \begin{itemize}
  \item[1.] Send the incoming particles to the
    \code{initHadronization} method to initialize the fragmentation.
  \item[2.] Ask the \code{getHadron} to administrate the creation of a
    new hadron.
  \item[3.] Re-invoke (2) while the remaining energy of the \String\ 
    is above a minimum value, given by the \code{Wmin2} method,
    otherwise ask the \code{final\-Two\-Hadron} method to produce the last
    two hadrons in the fragmentation process.
  \end{itemize}
\item{\bf\verb|initHadronization(const tPVector &);|}\\
  Is the method that create the \String\ object of the \lundfragh.
  Given the particles, coming from \code{Hadronize}, it has the
  responsibility to find the string type ({\it closed} or {\it open}
  string) and to call either the \code{initOpenString} or the
  \code{initClosedString} method that selects the correct initial
  particle to be sent to the \String\ constructor.  Besides that, it
  initializes the \lundfragh\ variables and prepares the iterative
  process by setting the correct rightmost and leftmost \endp s of the
  string.
\item{\bf\verb|getHadron();|}\\
  Administrates the creation of a new hadron. Its main task is :
  \begin{itemize}
  \item[1] Generate the new \endp\ in the current step.  For that
    \code{getHadron} makes use of the different \code{LundGenerator}s.
    Try to produce a new hadron joining the current and the last
    \endp.
  \item[2] If a solution is found then the full hadron kinematics is
    computed, otherwise it delegates to the \code{stepping} method the
    responsibility to find the correct breakup position that
    corresponds to a physically acceptable solution for the selected
    kinematics.
  \item[3] Invoke the \code{loopBack} method that properly updates the
    \String\ before starting a new step.
  \end{itemize}
\end{itemize}

\subsection{Outlook and comparison with \pythia\ version 6}
\label{sec:lundfrag:outlook}

The first implementation of the \pyth\ fragmentation modules provides
functionalities comparable to the default hadronization option of
\pythia\ version 6.1. However, some functionality, such as the
collapsing of low-mass strings into one or two particles, and the
treatment of so-called junction strings, is still missing.

The design and implementation of physics models for $p_t, z$ and
flavour generation have been completed, up to the level of the
\pythia6.1 functionality, for the \lundptgen\ and \lundzgen.  The
current version of the \lundflavourgen\ implements the whole meson
production model and the popcorn model for baryon production. An open
structure to handle the flavour generation in the fragmentation phase
has been provided for the foreseen implementation of the modified
popcorn model \cite{Patrik}.

Comparisons between the \pyth\ fragmentation modules and \pythia\ 
version 6 have been carried out at the level of test-case per class
and at the level of test-bench for the whole hadronization phase.
Excellent agreement between the two simulation results has been found.

\section{Bugs, lacking features and future plans}
\label{sec:bugs}

The current \pyth\ is a proof-of-concept version and it should
certainly not be assumed to be bug-free. In fact it should not even be
assumed to be able to produce anything particularly useful. But is
should be stable enough to start implementing real physics models. In
any case, anyone is more than welcome to play around with the code and
give suggestions or point out bugs to \code{leif@thep.lu.se}.

Most of what is missing in the current version of \pyth\ is physics
models, while most of the underlying structure where such models can
be implemented is in place. There are, however, some exceptions where
the current structure most likely needs to modified.

One such case is multiple interactions. Here the remnant handlers need
to be modified in order to allow for extraction of several partons
from a particle. The actual administration could be handled by a
specialized \pextract\ class, but one could also imagine introducing a
new group of step handlers.

For hard parton--parton scatterings, a new base class should be
introduced where it is possible to specify a general $2\rightarrow n$
matrix element.

\pyth\ is prepared for interfacing to the \SIUnits\ 
package\cite{SIUnits} for automatic compile-time checking of
dimensions of physical variables. To facilitate the transition to the
\SIUnits\ package, all dimensionful variables uses the \code{typedef}s
of \code{double} defined in the \classh{Units.h}{Units} file and all
dimensionful constants are multiplied with units as defined in the
\code{SystemOfUnits.h} file of \clhep. This means that variables are
handled as follows:
\begin{verbatim}
  Energy e = 2.5*GeV;
  Energy2 s = e*e;
  Energy m = sqrt(s);
\end{verbatim}

The user interface also needs to be improved. It is our hope that
someone else would take on the challenge of writing a cool graphical
user interface. This should not be too difficult as everything
important for a user interface is available through the \repository\ 
via the interface classes.

\end{document}